\documentclass[aps,prd,preprint,showpacs,floatfix,nofootinbib]{revtex4-1}

\usepackage{dcolumn}

\usepackage{amsmath, amsfonts, amssymb, mathrsfs, times, float, color, bm}
\usepackage{extarrows}
\usepackage{graphicx}
\usepackage{graphicx,epstopdf}
\usepackage{dcolumn}
\usepackage{bm}

\usepackage{exscale}
\usepackage{relsize}

\newcommand{\nn}{\nonumber}

\begin{document}

\title[]{Second Order Kerr-Newman Time Delay}

\author{G. He}
\author{W. Lin}
\email{wl@swjtu.edu.cn}

\affiliation{School of Physical Science and Technology,\\
 Southwest Jiaotong University, Chengdu 610031, China}

\begin{abstract}

The explicit form for the post-Newtonian gravitational time delay of light signals propagating on the equatorial plane of a Kerr-Newman black hole is derived. Based on the null geodesic in Kerr-Newman spacetime, we adopt the iterative method to calculate the time delay. Our result reduces to the previous formulation for Kerr black hole if we drop off the contribution from the electrical charge. Our time-delay formula for the Reissner-Nordstr\"{o}m geometry is different from the previous publication [Phys. Rev. D 69, 023002 (2004)], in which the largest second-order contribution to the time delay is missing.

\begin{description}
\item[PACS numbers] 98.62.Sb, 95.30.Sf, 04.70.Bw, 04.25.Nx
\end{description}

\end{abstract}

\maketitle

\section{Introduction}
Gravitational lensing attracts numerous attentions because of its extensive astronomical applications~\cite{BN1992,CHS2002,MM2003,WP2007,CLW2014,Leauthaud2015}. Within this geometrical effect, the gravitational time delay of electromagnetic waves acts as one of the most significant parts~\cite{Refsdal1964,BM2004,SPNJ2013,LT2013}, and has been investigated in detail. The pioneering work of obtaining the first-order gravitational delay of radar echo signals was carried out by Shapiro~\cite{Shapiro1964} to test general relativity~\cite{Shapiro1971}.
In later studies~\cite{RichterMatzner1983,Dymnikova1984,Dymnikova1986,GMBA1992,LW1997,CR2002,BM2004,Linwang2014}, the second-order relativistic corrections including mass- and spin-induced effects were considered to extend his result, owing to the rapid progresses made in observational accuracy~\cite{Perryman2001,Laskin2006,Lindegren2007,SN2009,Malbet2012,Malbet2014}. Especially, Richter and Matzner~\cite{RichterMatzner1983} used the Lagrangian function to investigate the second-order delay of signals caused by the spinning solar gravitational system, based on the parametrized post-linear metric.

The electrical charge of a lens system, serving as another type of gravitational source~\cite{ERT2002}, can produce a second-order relativistic correction on time delay, and deserves our attention as well~\cite{Punsly1998,PRT2000,ERT2002}. Nevertheless, time delay due to a charged gravitational source was not investigated until in last two decades. To our knowledge, the first calculation of the leading charge-induced contribution to time delay was accomplished in quasi-Minkowskian coordinates~\cite{Sereno2004}, based on Fermat's principle~\cite{SEF1992}. Keeton and Petters~\cite{KP2005} then developed their general lensing framework via the analysis of post-post-Newtonian formalism, in which the time delay between
positive- and negative-parity relativistic images in Reissner-Nordstr\"{o}m spacetime was derived.

Very recently, the Kerr-Newman black hole lensing, a more general scenario, started to receive its full considerations~\cite{Kraniotis2014,CS2015}, although some works had been done~\cite{ERT2002,BCJ2003,HP2006}. However, some significant basics in this situation, such as total signals delay, magnification relations, and the positions of the relativistic images, have not been investigated in the literatures. In this work, we derive the second-order time delay of light caused by Kerr-Newman black hole lensing in harmonic coordinates. Starting with the harmonic Kerr-Newman metric and null geodesic, we get the general integral form for coordinate time. Then we use the iterative method to obtain the analytical result, on the basis of the first-order equations of motion for light.

This paper is organized as follows. Section~\ref{derivation} presents the derivation for the explicit form of the second-order Kerr-Newman time delay. Section~\ref{Discussions} gives the discussions for the result. Summary is given in Section~\ref{Conclusions}. We use units $G = c = 1$ throughout.

\section{Second-order time delay of light caused by Kerr-Newman black hole} \label{derivation}

Consider the time delay for light in Kerr-Newman spacetime in the weak-field, small-angle, and thin lens approximation. Let $\{\bm{e}_1,~\bm{e}_2,~\bm{e}_3\}$ be the orthonormal basis of a three-dimensional Cartesian coordinate system. The metric of a Kerr-Newman black hole in harmonic coordinates $(X_0,~X_1,~X_2,~X_3)$, up to order $1/R^2$ within which the contributions of the angular momentum and electrical charge of the gravitational source appear, reads~\cite{LinJiang2015}
\begin{eqnarray}
&&g_{00}=-1-2\Phi-2\Phi^2-\frac{Q^2}{R^2}+O(1/R^3)~,  \label{g00KN} \\
&&g_{0i}=\zeta_i+O(1/R^3)~,  \label{g0iKN} \\
&&g_{ij}=(1-\Phi)^2\delta_{ij}+\frac{M^2-Q^2}{R^2}\frac{X_iX_j}{R^2}+O(1/R^3)~,  \label{gijKN}
\end{eqnarray}
where $i$ and $j$ run over the values $1,~2,~3$. $\delta_{ij}$ denotes Kronecker delta, and $\Phi\equiv-\frac{M}{R}$ represents Newtonian gravitational potential, with $\frac{X_1^2+X_2^2}{R^2+a^2}\!+\!\frac{X_3^2}{R^2}\!=\!1$ and $\mathbf{X} \! \cdot \! d\mathbf{X}\equiv X_1dX_1 \!+\! X_2dX_2 \!+\! X_3dX_3$.
$M$, $Q$, and $\bm{J} (=\!J\bm{e}_3)$ are the rest mass, electrical charge and angular momentum vector along the positive $X_3-$axis of the source, respectively.
$\bm{\zeta}\!\equiv\!\frac{2aM}{R^3}\left(\bm{X} \!\times\! \bm{e_3}\right)=(\zeta_1,~\zeta_2,~0)$, and $a \!\equiv\! \frac{J}{M}$ is the angular momentum per mass. The relation $M^2\geq a^2+Q^2$ is assumed to avoid naked singularity for the black hole.

\begin{figure*}[t]
\begin{center}
  \includegraphics[width=13cm]{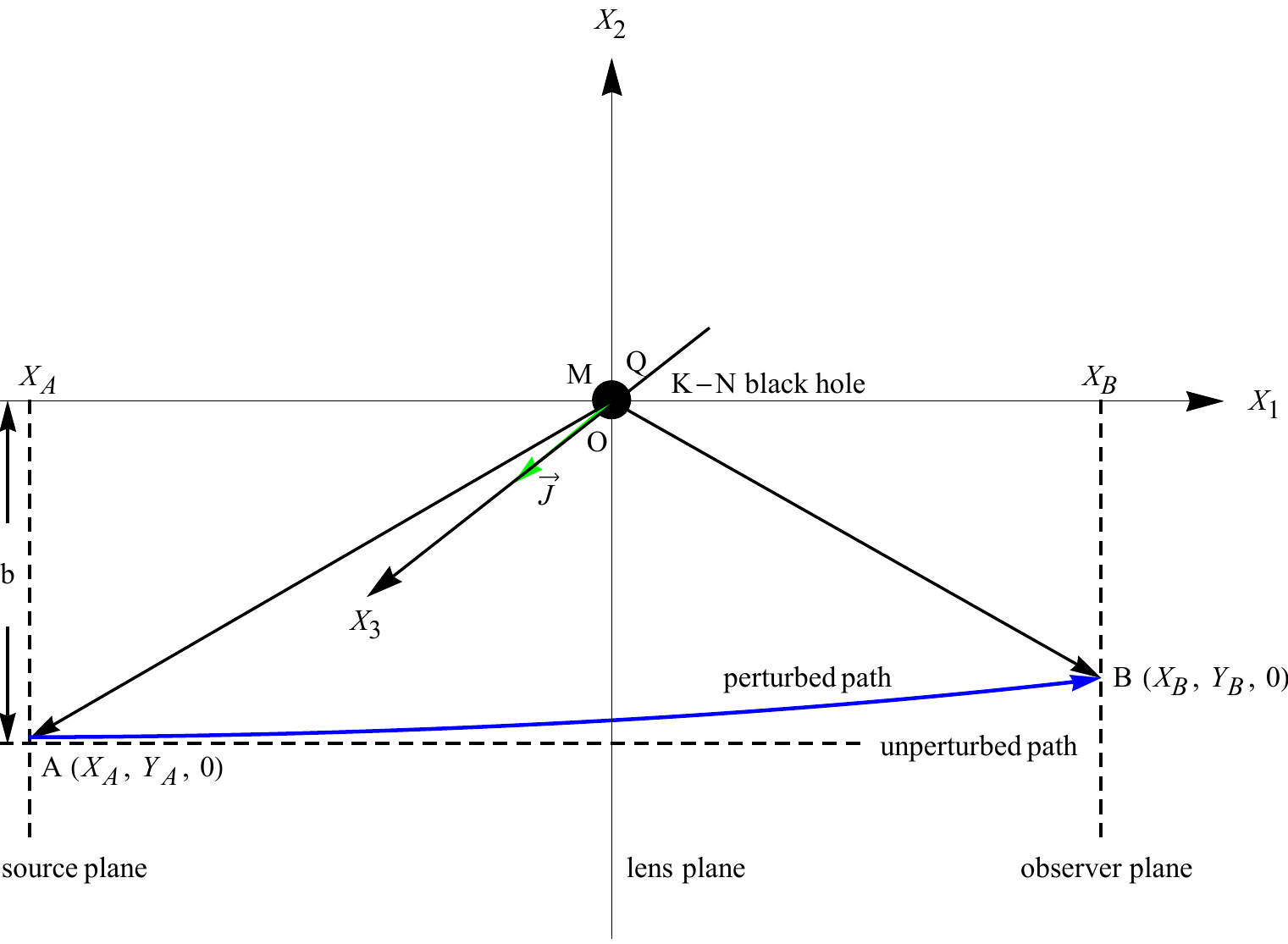}
  \caption{Schematic diagram for gravitational time delay of light signals caused by Kerr-Newman black hole. The locations of the light source (denoted by $A$ on the source plane) and observer (denoted by $B$ on the observer plane) are respectively supposed to be $(X_A,~Y_A,~0)$ and $(X_B,~Y_B,~0)$, where $Y_A\approx -b$, $|X_A|\gg b~(X_A<0)$, and $X_B\gg b$, with $b$ being the impact parameter. We assume that light takes the prograde motion relative to the rotation $\bm{J} (=\!J\bm{e}_3)$ of gravitational source. The gravitational deflection is greatly exaggerated to distinguish the solid blue line, which represents the perturbed propagation path of light, from the dashed horizontal line which denotes the unperturbed one. Notice that here the source and lens planes are perpendicular to the asymptote of the incoming ray, which is different from the geometry of the standard lensing formalism where they are usually perpendicular to the axis through the observation event. }    \label{Figure1}
\end{center}
\end{figure*}

The general form of null curve is expressed as
\begin{equation}
0=ds^2=g_{\mu\nu}dx^\mu dx^\nu~,    \label{geodesic1}
\end{equation}
where the indices $\mu,~\nu$ run over the values $0,~1,~2,~3$. For simplicity, we assume light propagation is confined to the equatorial plane of the black hole, i.e., $X_3=\frac{\partial }{\partial X_3}=0$. The corresponding schematic model for the propagation of light signals
is shown in Fig.~\ref{Figure1}.
Thus, Eq.~\eqref{geodesic1} can be written as
\begin{equation}
0=g_{00}\,dX_0^2+g_{11}\,dX_1^2+g_{22}\,dX_2^2+2\,g_{01}\,dX_0dX_1+2\,g_{02}\,dX_0dX_2+2\,g_{12}\,dX_1dX_2 ~.    \label{geodesic2}
\end{equation}
From this equation, we have
\begin{equation}
\frac{dX_0}{dX_1}=\frac{-E+\sqrt{E^2-4F}}{2} ~,    \label{geodesic4}
\end{equation}
where
\begin{eqnarray}
&&E=2\left(\frac{g_{01}}{g_{00}}+\frac{g_{02}}{g_{00}}\frac{dX_2}{dX_1}\right) ~,  \label{E} \\
&&F=\frac{g_{11}}{g_{00}}+\frac{2g_{12}}{g_{00}}\frac{dX_2}{dX_1}+\frac{g_{22}}{g_{00}}\left(\frac{dX_2}{dX_1}\right)^2 ~.  \label{F}
\end{eqnarray}
Here the other solution which is not physical has been omitted, since $F=-1+O(1/R)<0$ under the weak-field and small-angle hypotheses. Notice that $\frac{dX_2}{dX_1}$ is related to the gravitational deflection angle $\alpha$ of light by $\left.\alpha \approx \arctan{\frac{dX_2}{dX_1}}\right|_{X_1=X_B}-\left.\arctan{\frac{dX_2}{dX_1}}\right|_{X_1=X_A} \approx \left.\arctan{\frac{dX_2}{dX_1}}\right|_{X_1=X_B}$.

In order to obtain the analytical coordinate time $X_0$, we perform an indefinite integral over $X_1$ for Eq.~\eqref{geodesic4}
\begin{eqnarray}
X_0=\mathlarger{\int}\frac{-E+\sqrt{E^2-4F}}{2}dX_1
   =\mathlarger{\int}\left(\sqrt{-\frac{g_{11}}{g_{00}}+\left(\frac{dX_2}{dX_1}\right)^2}+g_{01}\right)dX_1~.  \label{Integral-1}
\end{eqnarray}
In the second equality, we have dropped off the third and higher order terms, such as $\frac{g_{02}}{g_{00}}\frac{dX_2}{dX_1}$ and $\frac{g_{12}}{g_{00}}\frac{dX_2}{dX_1}$, since we only consider the time delay up to second order.

Now we derive the explicit form of $dX_2/dX_1$ up to first order. We start with the equations of motion for light in the field of a Schwarzschild black hole, which up to the first post-Minkowskian order (1PM) can be written as~\cite{LinHe2015}
\begin{eqnarray}
&&0=\ddot{X_0}+\frac{2\dot{X_0}\dot{X_1}X_1M}{R^3}+O\left(M^2\right)  ~,   \label{motonEQ-t}   \\
&&0=\ddot{X_1}+\frac{\left(\dot{X}_0^2-\dot{X}_1^2\right) X_1 M}{R^3}+O\left(M^2\right)  ~,  \label{motonEQ-x}   \\
&&0=\ddot{X_2}+\frac{\left(\dot{X}_0^2+\dot{X}_1^2\right) X_2 M}{R^3}+O\left(M^2\right)  ~,  \label{motonEQ-y}
\end{eqnarray}
where a dot denotes the derivative with respect to the affine parameter $p$ which describes the trajectory of light~\cite{Weinberg1972,WuckSperh2004}. $R$ is reduced to $R=\sqrt{X_1^2+X_2^2}$ up to 1PM order. Here we set $p$ to be $X_1$, as done in Ref.~\cite{WuckSperh2004}.

Combining Eqs.~\eqref{motonEQ-t} - \eqref{motonEQ-x} with the boundary conditions $\dot{X_0}|_{X_1\rightarrow-\infty}~(\approx \dot{X_0}|_{X_1\rightarrow X_A})=1$
and $\dot{X_1}|_{X_1\rightarrow-\infty}~(\approx \dot{X_1}|_{X_1\rightarrow X_A})=1$, we obtain the zero-order values for $\dot{X_0}$ and $\dot{X_1}$ as
\begin{eqnarray}
&&\dot{X_0}=1+O\left(M\right)  ~,  \label{motonEQ-t-2}   \\
&&\dot{X_1}=1+O\left(M\right)  ~.  \label{motonEQ-x-2}
\end{eqnarray}
Substituting Eqs.~\eqref{motonEQ-t-2} - \eqref{motonEQ-x-2} into Eq.~\eqref{motonEQ-y}, and integrating $X_1$ for Eq. \eqref{motonEQ-y}, we can obtain the first-order analytical form for $\dot{X_2}$ as follow
\begin{equation}
\frac{dX_2}{dX_1}=\frac{2M}{b}\left(1+\frac{X_1}{\sqrt{X_1^2+b^2}}\right)+O\left(M^2\right) ~,    \label{motonEQ-y-2}
\end{equation}
where the boundary conditions $\dot{X_2}|_{X_1\rightarrow-\infty}~(\approx \dot{X_2}|_{X_1\rightarrow X_A})=0$ and $X_2|_{X_1\rightarrow-\infty}~(\approx X_2|_{X_1\rightarrow X_A})=-b$ (zero-order approximation) have been used. When $X_1=X_B$~, one can see that the gravitational deflection angle $\alpha$ up to first order is $\alpha \approx 4M/b$~.

Plugging Eqs.~\eqref{g00KN} - \eqref{gijKN} and \eqref{motonEQ-y-2} into Eq.~\eqref{Integral-1}, we have
\begin{eqnarray}
\nn&& X_0=\mathlarger{\int}\left[\sqrt{\frac{1+\frac{2M}{R}+\frac{M^2}{R^2}+\frac{(M^2-Q^2)\,X_1^2}{R^4}}{1-\frac{2M}{R}+\frac{2M^2+Q^2}{R^2}}
+\frac{4M^2}{b^2}\left(1+\frac{X_1}{\sqrt{X_1^2+b^2}}\right)^2}+\frac{2\,aM X_2}{R^3}\right]dX_1¡¡  \\
\nn&&\hspace*{12pt} \!\!\approx \!\! \mathlarger{\int}\!\left[\sqrt{1\!+\!\frac{4M}{R}\!+\!\frac{7M^2\!-\!Q^2}{R^2}\!+\!\frac{(M^2\!-\!Q^2)\,X_1^2}{R^4}
\!+\!\frac{4M^2}{b^2}\left(1+\frac{X_1}{\sqrt{X_1^2+b^2}}\right)^2}\!+\!\frac{2\,aM X_2}{R^3}\right]\! dX_1  \\
\nn&&\hspace*{12pt} \!\!\approx \!\! \mathlarger{\int}\!\!\left[1\!+\!\frac{2M}{R}\!-\!\frac{2M^2}{R^2}\!+\!\frac{7M^2\!-\!Q^2}{2R^2}\!+\!\frac{(M^2\!-\!Q^2)X_1^2}{2R^4}
\!+\!\frac{2M^2}{b^2}\!\left(\!1\!+\!\frac{X_1}{\sqrt{X_1^2\!+\!b^2}}\!\right)^{\hspace*{-2pt}2}\!\!+\!\frac{2aM X_2}{R^3}\!\right] \! dX_1~.  \\ ~~~~\label{Integral-2}
\end{eqnarray}

On the other hand, integrating Eq.~\eqref{motonEQ-y-2} over $X_1$ also results in
\begin{equation}
X_2=-b\left[1-\frac{2M\left(X_1+\sqrt{X_1^2+b^2}\right)}{b^2}\right]+O(M^2)~,~~\left(\frac{2M(X_1+\sqrt{X_1^2+b^2})}{b^2}\ll1\right)~,~ \label{y}
\end{equation}
where the integral constant has been determined by $\left.X_2\right|_{X_1\rightarrow X_A}\approx -b$. Substituting Eq.~\eqref{y} into Eq.~\eqref{Integral-2}, we have
\begin{eqnarray}
\nn&&X_0=\!\mathlarger{\int}\Bigg\{1+\frac{2\,M}{\sqrt{X_1^2+b^2}\sqrt{1-4\,M\left(X_1+\sqrt{X_1^2+b^2}\right)/\left(X_1^2+b^2\right)}}-\frac{2M^2}{X_1^2+b^2}
-\frac{2\,aMb}{(X_1^2+b^2)^{\frac{3}{2}}}    \\
\nn&&\hspace*{30pt}+\frac{1}{2}\left[\frac{7M^2-Q^2}{X_1^2+b^2}+\frac{\left(M^2-Q^2\right)X_1^2}{\left(X_1^2+b^2\right)^2}
+\frac{4M^2}{b^2}\left(1+\frac{X_1}{\sqrt{X_1^2+b^2}}\right)^2\right]\Bigg\}dX_1   \\
\nn&&\hspace*{17pt} \approx \!\mathlarger{\int}\left[1+\frac{2M}{\sqrt{X_1^2+b^2}}+\frac{4M^2\left(X_1+\sqrt{X_1^2+b^2}\right)}{(X_1^2+b^2)^{\frac{3}{2}}}
-\frac{2M^2}{X_1^2+b^2}-\frac{2\,aMb}{(X_1^2+b^2)^{\frac{3}{2}}}+\frac{7M^2-Q^2}{2(X_1^2+b^2)}  \right.  \\
\nn&&\hspace*{30pt}\left.+\frac{(M^2-Q^2)X_1^2}{2\left(X_1^2+b^2\right)^2}+\frac{2M^2}{b^2}\left(1+\frac{X_1}{\sqrt{X_1^2+b^2}}\right)^2 \,\right] dX_1   \\
\nn&&\hspace*{17pt} =X_1+2M \,ln\left(X_1+\sqrt{X_1^2+b^2}\,\right)+4M^2\left(\!-\frac{1}{\sqrt{X_1^2+b^2}}+\frac{1}{b}\arctan{\frac{X_1}{b}}\!\right)
-\frac{2\,aM X_1}{b\,\sqrt{X_1^2+b^2}} \\
\nn&&\hspace*{30pt} +\frac{2M^2}{b^2}\!\left(2X_1+2\sqrt{X_1^2+b^2}-b\arctan{\frac{X_1}{b}}\right)
\!+\!\frac{M^2\!-\!Q^2}{2}\!\left[-\frac{X_1}{2\left(X_1^2+b^2\right)}\!+\!\frac{1}{2\,b}\arctan{\frac{X_1}{b}}\right]      \\
&&\hspace*{30pt}+\frac{3M^2-Q^2}{2\,b}\arctan{\frac{X_1}{b}}+C  ~,    \label{Integral-3}
\end{eqnarray}
where $C$ denotes the integral constant, and we have dropped off the third and higher order terms in the derivation. The explicit form of the coordinate time up to second order for light propagating from $A~(X_A,~Y_A,~0)$ to $B~(X_B,~Y_B,~0)$ in Fig.~\ref{Figure1} can be written as
\begin{eqnarray}
\nn&&X_0\,(X_B, X_A)=(X_B-\!X_A)+2M\,ln\!\left(\frac{\sqrt{X_B^2+b^2}+X_B}{\sqrt{X_A^2+b^2}+X_A}\!\right)
\!+\!4M^2\!\left(\frac{1}{\sqrt{X_A^2+b^2}}-\frac{1}{\sqrt{X_B^2+b^2}}\right)    \\
\nn&&\hspace*{40pt}+\frac{4M^2}{b^2}\left(X_B-X_A+\sqrt{X_B^2+b^2}-\sqrt{X_A^2+b^2}\,\right)+\frac{M^2-Q^2}{4}\left(\frac{X_A}{X_A^2+b^2}-\frac{X_B}{X_B^2+b^2}\right)   \\
&&\hspace*{40pt}+\frac{3\left(5M^2\!-\!Q^2\right)}{4b}\left(\!\arctan{\frac{X_B}{b}}\!-\!\arctan{\frac{X_A}{b}}\!\right)
\!+\!\frac{2aM}{b}\left(\frac{X_A}{\sqrt{X_A^2+b^2}}\!-\!\frac{X_B}{\sqrt{X_B^2+b^2}} \right)~.      \label{Integral-4}
\end{eqnarray}
Considering $|X_A|\gg b$ and $X_B \gg b$, finally we obtain
\begin{eqnarray}
\nn&&X_0\,(X_B, X_A)=(X_B-\!X_A)\!+2M\,ln\!\left(\frac{\sqrt{X_B^2\!+\!b^2}\!+\!X_B}{\sqrt{X_A^2\!+\!b^2}\!+\!X_A}\right)\!+\!\frac{8M^2X_B}{b^2}
\!+\!\frac{3\pi\left(5M^2\!-\!Q^2\right)}{4b}-\frac{4aM}{b} ~, \\
&& \label{Integral-5}
\end{eqnarray}
where the second-order terms with the factor $\frac{1}{X_A}$ or $\frac{1}{X_B}$, such as $\frac{4M^2}{X_A}$ and $\frac{Q^2}{4X_B}$, have been neglected, since they are much smaller than the second-order terms containing the factor $\frac{1}{b}$~~(e.g., the term $\frac{15\pi M^2}{4b}$).

\section{Discussions} \label{Discussions}

The leading term on the right-hand side of Eq.~\eqref{Integral-5} represents the geometrical time for light travelling in straight lines. The second term is the first-order relativistic correction for the time delay due to the gravitational source, which is consistent with Shapiro's primary result in Refs.~\cite{Shapiro1964,Shapiro1971} and the later derivation in Ref.~\cite{KopeiSch1999}. The rest three terms are second order, and reduce to the results in Ref.~\cite{RichterMatzner1983} when we drop off the contribution from the charge ($Q=0$). The detailed comparisons with other works~\cite{Shapiro1971,KopeiSch1999,RichterMatzner1983} are given in Appendix A.

In order to compare with the Reissner-Nordstr\"{o}m time delay of light in the previous work~\cite{Sereno2004}, we consider the case of $X_A\approx-X_B=-X_{max}~(X_{max}>0)$ and $a=0$. In this case, we have
\begin{eqnarray}
X_0(X_{max}, -X_{max})= 2X_{max}+4M\,ln\frac{2X_{max}}{b}+\frac{8M^2X_{max}}{b^2}+\frac{3\pi\left(5M^2-Q^2\right)}{4b}~.~~~~ \label{Integral-10}
\end{eqnarray}
We find that the charge-induced contribution on the right-hand side of Eq.~\eqref{Integral-10} coincides with the previous result, however, the term $\frac{8M^2X_{max}}{b^2}$ in Eq.~\eqref{Integral-10} is missing in Ref.~\cite{Sereno2004} (see Eq. (12) therein). This term, which is caused by $\frac{g_{22}}{g_{00}}\left(\frac{dX_2}{dX_1}\right)^{\hspace*{-1pt} 2}$ in Eq.~\eqref{F}, is the largest second-order relativistic correction on the delay, and can not be neglected. In fact, this term has also been obtained via Euler-Lagrange method~\cite{RichterMatzner1983}.

\section{Summary} \label{Conclusions}
We have derived the analytical formula for the total second-order gravitational time delay in the Kerr-Newman geometry based on the equations of motion for light. In the limit of no charge, our result reduces to the formula obtained via Euler-Lagrange method~\cite{RichterMatzner1983}, and agrees with Shapiro's pioneering work~\cite{Shapiro1964,Shapiro1971} and the later derivation given in Ref.~\cite{KopeiSch1999}. The charge-induced contribution to the time delay is in agreement with that in Ref.~\cite{Sereno2004}. On the other hand, we find that the largest second-order contribution on gravitational delay of light, which has been derived in Ref.~\cite{RichterMatzner1983}, is missing in Ref.~\cite{Sereno2004}.

\section*{ACKNOWLEDGEMENT}
This work was supported in part by the National Natural Science Foundation of China (Grant No. 11547311), the National Basic Research Program of China (Grant No. 2013CB328904) and the Fundamental Research Funds for the Central Universities (No. 2682014ZT32)

\appendix

\section{Comparisons with the results in previous works} \label{appendix}

The primary work for the one-way gravitational time retardation (1PM order) in harmonic coordinates reads~\cite{Shapiro1971}
\begin{eqnarray}
\Delta t= \frac{2r_0}{c}~ln\left(\frac{r_e+r_p+R}{r_e+r_p-R}\right)~, \label{A1}
\end{eqnarray}
which is rewritten in our notations as
\begin{eqnarray}
\Delta t=2M~ln\left(\frac{\sqrt{X_B^2+b^2}+\sqrt{X_A^2+b^2}+|X_A|+X_B}{\sqrt{X_B^2+b^2}+\sqrt{X_A^2+b^2}-|X_A|-X_B}\right)~. \label{A2}
\end{eqnarray}
To compare Eq.~\eqref{A2} with the second term on the right-hand side of Eq.~\eqref{Integral-5}, we just need to prove
\begin{eqnarray}
\frac{\sqrt{X_B^2+b^2}+\sqrt{X_A^2+b^2}+|X_A|+X_B}{\sqrt{X_B^2+b^2}+\sqrt{X_A^2+b^2}-|X_A|-X_B}-\frac{\sqrt{X_B^2+b^2}+X_B}{\sqrt{X_A^2+b^2}+X_A}=0 ~, \label{A3}
\end{eqnarray}
which is not difficult to be verified, with $X_A=-|X_A|$.

For a stationary point mass (with a mass $m_1$) as the gravitational system, the first post-Minkowskian time delay derived in Ref.~\cite{KopeiSch1999} (see,~Eq.~(51) and Fig.~4 therein) is
\begin{eqnarray}
\Delta(t,~t_0)=2m_1~ln\left(\frac{r_{01}-\vec{k}\cdot\vec{r}_{01}}{r_1-\vec{k}\cdot\vec{r}_1} \right)~.   \label{A4}
\end{eqnarray}
Notice that in our notations $m_1=M$,~$\vec{k}=-\vec{e}_1$,~$r_{01}=|\vec{r}_{01}|$,~$r_1=|\vec{r}_1|$, $\vec{r}_{01}\approx X_B\vec{e}_1-b\vec{e}_2$ and
$\vec{r}_1\approx-|X_A|\vec{e}_1-b\vec{e}_2$, since the light source and observer in Fig.~4 of Ref.~\cite{KopeiSch1999} are located at our observer and source planes, respectively. Therefore, Eq.~\eqref{A4} is simplified to the same form as that in Eq.~\eqref{Integral-5}
\begin{eqnarray}
\Delta(t,~t_0)=2M~ln\left(\frac{\sqrt{X_B^2+b^2}+X_B}{\sqrt{X_A^2+b^2}+X_A} \right)~.   \label{A5}
\end{eqnarray}

We then consider the results presented in Ref.~\cite{RichterMatzner1983}, which are calculated in the parameterized post-Newtonian (PPN) coordinates system $(x,~y,~z)$. Notice that PPN coordinates are related to harmonic coordinates by $r=\sqrt{x^2+y^2+z^2}=R+M$, where $r$ and $R$ are the radius variables in standard and harmonic coordinates, respectively.

At first, we discuss the special case that the locations of the the ray transmitter ($x_T,~y_T,~0$) and reflector ($x_R,~y_R,~0$) are approximately symmetrical with respect to
the deflector plane (i.e., $r\equiv r_R\simeq r_T,~y_T\simeq b>0,~x_T<0,~x_R>0$) in Ref.~\cite{RichterMatzner1983}. The one-way time delay for a radar signal just grazing the Sun on the equatorial plane is given as follow (see Eq.~(3.24) therein)
\begin{eqnarray}
\tau_{TR}=2r-2M+\frac{8M^2r}{y_T^2}+\frac{M^2}{r}+2M~ln\frac{4r^2}{y_T^2}-\frac{2M^2}{r}~ln\frac{4r^2}{y_T^2}+\frac{15M^2}{2y_T}\arctan{\frac{r}{y_T}}\pm\frac{4J}{y_T}~,
~~~~~~~~ \label{A6}
\end{eqnarray}
which has been restricted to general relativity and the case of no quadrupole moment. $M$ and $J$ are the rest mass and angular momentum of the Sun, respectively, and the radius of it has been replaced by $y_T$ (see Eq.~(3.18) therein). Eq.~\eqref{A6} is expressed in our notations as
\begin{eqnarray}
\nn&&\tau_{AB}\approx 2R+4M~ln\frac{2R}{b}+\frac{8M^2R}{b^2}+\frac{15M^2}{2b}\arctan{\frac{R}{b}}\pm\frac{4J}{b}  \\
&&\hspace*{23pt}\approx 2X_{max}+4M~ln\frac{2X_{max}}{b}+\frac{8M^2X_{max}}{b^2}+\frac{15\pi M^2}{4b}\pm\frac{4J}{b}~,~~(X_{max}=X_B\approx -X_A)~,~~~~~~~~\label{A7}
\end{eqnarray}
where we have neglected of the third and higher order terms and the second-order terms including the factor $\frac{1}{X_{max}}$ for $b\ll X_{max}$, as done in Section~\ref{derivation}.
Note that the term $\frac{a^2}{X_{max}}~(\leq\frac{M^2}{X_{max}})$, produced by $R=\sqrt{X_1^2+X_2^2-a^2}\approx \sqrt{X_1^2+b^2}-a^2/(2\sqrt{X_1^2+b^2})$, has also been omitted. It can be seen that Eq.~\eqref{A7} is consistent with Eq.~\eqref{Integral-5} for $Q=0$ since $ln\!\left(\frac{\sqrt{X_{max}^2+b^2}+X_{max}}{\sqrt{X_{max}^2+b^2}-X_{max}}\right)\approx 2\,ln\frac{2X_{max}}{b}$.

Now we consider the consistency between our result and the formula in Ref.~\cite{RichterMatzner1983} for a general case, i.e., $r_R\neq r_T$. We focus on the pure second-order contributions on the one-way time delay on the right-hand side of Eq.(3.14) therein, which are written as
\begin{eqnarray}
\nn&&\Delta\tau_{TR}\!=\!(x_R\!-\!x_T)\!\left\{\!\frac{2M^2}{b^2}\!\!\left[\!\left(\!1\!-\!\frac{x_T}{r_T}\right)^2\!\!+\!\frac{2(x_R x_T\!-\!b^2)}{r_R r_T}\!\right]
\!+\!\frac{M^2}{2r_T^2}\!\right\}\!+\!\frac{15 M^2}{4b}\!\left(\!\arctan{\frac{x_R}{b}}\!-\!\arctan{\frac{x_T}{b}}\right)  \\
&&\hspace*{40pt}-\,\frac{2M^2}{r_T}ln\frac{x_R+r_R}{x_T+r_T}+\frac{4M^2r_T}{b^2}\left(1-\frac{r_R}{r_T}\right)\left(\frac{x_T}{r_T}+\frac{x_R}{r_R}-1\right)
\pm\frac{2J}{b}\left(\frac{x_R}{r_R}-\frac{x_T}{r_T}\right)~,~~~~\label{A8}
\end{eqnarray}
where $y_T\!\approx b\left[1+O\left(M\right)\right]$ has been used. Notice that for the delay up to second order we have $r_T=r_A=R_A+M$,~$r_R=r_B=R_B+M$,~$x_T=x_A\approx-r_A\approx X_A-M$,
and $x_R=x_B\approx r_B\approx X_B+M$. Therefore, Eq.~\eqref{A8} can be written in our notations as the same form in Eq.~\eqref{Integral-5} for the case of $Q=0$
\begin{eqnarray}
&&\Delta\tau_{AB}\approx \frac{8M^2X_B}{b^2}+\frac{15\pi M^2}{4b}\pm\frac{4J}{b} ~, ~~~~\label{A9}
\end{eqnarray}
where we have dropped off the third and higher order terms, as well as all the second-order terms containing $\frac{1}{X_A}$ or $\frac{1}{X_B}$,
which are much less than the second-order terms containing $\frac{1}{b}$ (e.g., the term $\frac{15\pi M^2}{4b}$).

\end{document}